\documentclass[manuscript,onecolumn,preprintnumbers,superscriptaddress]{revtex4}
\usepackage{amssymb}
\usepackage{amsmath}
\usepackage{graphicx}
\usepackage{dcolumn}
\usepackage{bm}
\usepackage{multirow}

\begin{document}

\title{Multiferroic hexagonal ferrites (h-RFeO$_3$, R=Y, Dy-Lu): an experimental review}

\author{Xiaoshan Xu}
\affiliation{Department of Physics and Astronomy, University of Nebraska-Lincoln, Lincoln, NE 6858, USA}

\author{Wenbin Wang}
\affiliation{Department of Physics, Fudan University, Shanghai 200433, China}

\date{\today}

\begin{abstract}
  Hexagonal ferrites (h-RFeO$_3$, R=Y, Dy-Lu) have recently been identified as a new family of multiferroic complex oxides.
 The coexisting spontaneous electric and magnetic polarizations make h-RFeO$_3$ rare-case ferroelectric ferromagnets at low temperature.
 Plus the room-temperature multiferroicity and predicted magnetoelectric effect, h-RFeO$_3$ are promising materials for multiferroic applications.
 Here we review the structural, ferroelectric, magnetic, and magnetoelectric properties of h-RFeO$_3$.
 The thin film growth is also discussed because it is critical in making high quality single crystalline materials for studying intrinsic properties.
\end{abstract}


\maketitle

\tableofcontents


	\section{Introduction}
 In the quest for energy-efficient and compact materials for information processing and storage, magnetoelectric multiferroics stand out as promising candidates.\cite{Khomskii2009,Spaldin2010}
 In particular, iron based oxide materials are an important category because of the strong magnetic interactions of the Fe sites.
 For example, BiFeO$_3$ is arguably the most studied multiferroic material due to the coexistence of ferroelectricity and antiferromagnetism above room temperature.\cite{Catalan2009}
 BaO-Fe$_2$O$_3$-MeO ferrites of hexagonal structure (Me=divalent ion such as Co, Ni, and Zn; Ba can be substituted by Sr) exhibit room-temperature magnetoelectric effect.\cite{Pullar2012,Kitagawa2010,Soda2011}
 LuFe$_2$O$_4$ is a unique ferrimagnetic material with antiferroelectricity originated from charge order.\cite{Christianson2008,Groot2012}
 Spontaneous electric and magnetic polarizations occur simultaneously in hexagonal ferrites h-RFeO$_3$ (R=Y, Dy-Lu) at low temperature;\cite{Akbashev2011,Jeong2012,Jeong2012b,Iida2012,Wang2013}
 the antiferromagnetism and ferroelectricity is demonstrated to persists above room temperature.\cite{Wang2013}

 In this review, we focus on hexagonal ferrites h-RFeO$_3$, which is isomorphic to hexagonal RMnO$_3$.
 Although the stable structure of free-standing RFeO$_3$ are orthorhombic, the hexagonal structure can be stabilized using various methods.\cite{Yamaguchi1991,Mizoguchi1996,Nagashio2002,Bossak2004,Nagashio2006,Kuribayashi2008,Li2008,Kumar2008,Kumar2009,Hosokawa2009,Magome2010,Akbashev2011,Hosokawa2011,Pavlov2012,Akbashev2012,Akbashev2012b,Iida2012,Jeong2012,Jeong2012b,Roddatis2013,Wang2013,Masuno2013,Moyer2014}
 Similar to hexagonal RMnO$_3$ (space group P6$_3$cm), h-RFeO$_3$ exhibit ferroelectricity and antiferromagnetism.\cite{Jeong2012,Jeong2012b,Wang2013,Fennie2005,Munoz2000}
 What distinguishes h-RFeO$_3$ from RMnO$_3$ is the low temperature weak ferromagnetism and the higher magnetic transition temperature due to the stronger exchange interaction between the Fe$^{3+}$ sites; the spontaneous magnetization can be magnified by the magnetic R$^{3+}$ sites.\cite{Jeong2012,Iida2012}
 The theoretically predicted reversal of magnetization using an electric field is another intriguing topic in h-RFeO$_3$.\cite{Das2014}
 We will review the experimental aspects of the h-RFeO$_3$, including their structural, ferroelectric, magnetic, and magnetoelectric properties.
 Since h-RFeO$_3$ is a metastable state of RFeO$_3$, the growth of materials involve special treatment; we will discuss the stability of h-RFeO$_3$ and emphasize especially on the thin film growth.
 In addition to a summary of the existing experimental work, we also included some new analysis, representing the current understanding of the authors.

\subsection{Structure}

\begin{figure}[ht]
\centerline{
\includegraphics[width = 4 in]{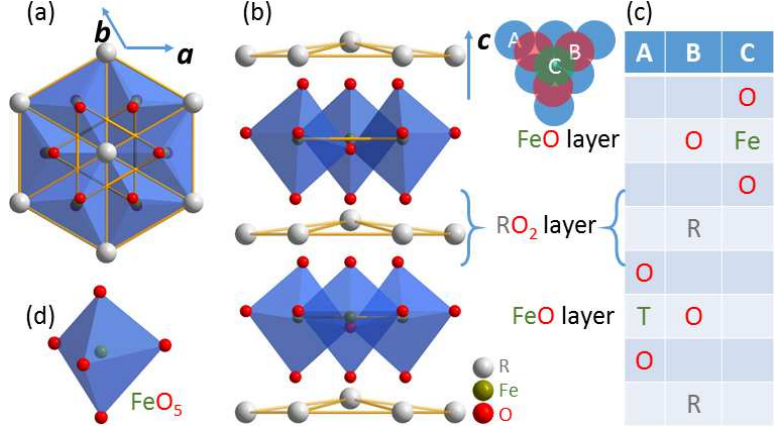}
}
\caption{ (Color online)
 Structure of the h-RFeO$_3$. (a) Viewed along the $c$-axis. (b) Viewed along the $a$-axis. (c) Viewed in terms of the ABC stacking. (d) FeO$_5$ as the local environment of Fe atoms.
}
\label{Fig_RFeO3_structure}
\end{figure}

 The structure of the h-RFeO$_3$ at room temperature belongs to a P6$_3$cm space group with a six-fold rotational symmetry, as shown in Fig. \ref{Fig_RFeO3_structure}.\cite{Magome2010}
 The unit cell can be divided into four layers: two RO$_2$ layers and two FeO layers. 
 The arrangements of the atoms follow roughly the ABC hexagonal stacking [Fig. 4(c)]. 
 The Fe atoms occupy the two dimensional triangular lattice in the FeO layer. 
 Every Fe atom is surrounded by five oxygen atoms (three in the same FeO layer and one above and one below the FeO layer), forming a FeO$_5$ trigonal bipyramid [Fig. \ref{Fig_RFeO3_structure}(d)]. 
 Each R atom is surrounded by eight oxygen atoms (six in the same RO$_2$ layer, one above and one below the RO$_2$ layer), forming a RO$_8$ local environment. 
 Note that the FeO$_5$ is slightly rotated along the [120] crystal axis; 
 this rotation causes the broken inversion symmetry of the h-RFeO$_3$ structure, allowing for the ferroelectricity.\cite{Fennie2005,Das2014}

\subsection{Ferroelectricity}

\label{marker:ferroelectricity}

\begin{figure}[ht]
\centerline{
\includegraphics[width = 4 in]{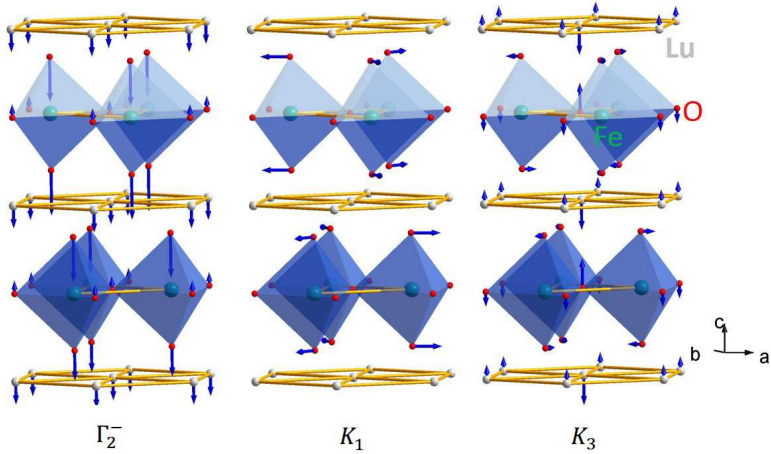}
}
\caption{ (Color online)
 The illustration of the three phonon modes ($\Gamma_2^-$, K$_1$ and K$_3$) related to the P6$_3$/mmc to P6$_3$cm structural transition.
 We use the coordinate system of P6$_3$cm structure here (and throughout the review) for the $a$, $b$ and $c$ axis.
 The rods connecting atoms are not to indicate chemical bonds, but to highlight the structural symmetry.
}
\label{Fig_phononmodes}
\end{figure}

 Hexagonal transition metal oxides have a P6$_3$/mmc structure at high temperature (T$_C\approx$1000 K, Fig. \ref{Fig_phononmodes}) and a P6$_3$cm structure at room temperature (Fig. \ref{Fig_RFeO3_structure}). 
 In the transition from the P6$_3$/mmc structure to the P6$_3$cm structure, three structural changes occur, which can be viewed as the frozen phonon modes $\Gamma_2^-$, K$_1$ and K$_3$, as shown in Fig. \ref{Fig_phononmodes}. The $\Gamma_2^-$ mode corresponds to the uneven displacement of the atoms along the $c$-axis; this generates the spontaneous electric polarization (ferroelectricity). 
 The K$_3$ mode corresponds to a collective rotation of the FeO$_5$ trigonal bipyramids; it turns out to be the driving force for the structural transition that causes the none-zero displacement of the $\Gamma_2^-$ modes.\cite{Fennie2005,Das2014}
 As shown in Fig. \ref{Fig_phononmodes}, the direction of the FeO$_5$ rotation actually decides whether the majority of the R atoms are above or below the minority in the RO$_2$ layers, which in turn decides the direction of the spontaneous electric polarization.

\subsection{Magnetism}

\label{marker:intromagnetism}

\begin{figure}[ht]
\centerline{
\includegraphics[width = 2.5 in]{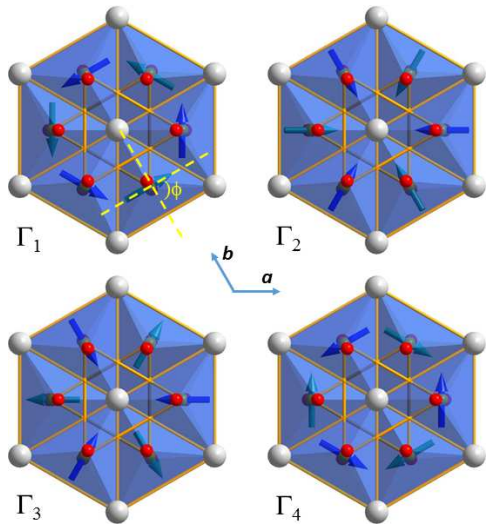}
}
\caption{ (Color online)
 Four ($\Gamma_1$ to $\Gamma_4$) independent 120-degree antiferromagnetic orders of the spins on the Fe sites in h-RFeO$_3$. 
 The blue and green arrows represent the spins in $z=0$ and $z=\frac{c}{2}$ FeO layers viewed along the $c$-axis. 
 The four spin structures come from the combination of two $\phi$ angles and two relative alignments of the spins between the two FeO layers (parallel or antiparallel). 
 Only $\Gamma_2$ allows for spontaneous magnetic polarizations.
 }
\label{Fig_spinstructures}
\end{figure}

 The Fe site in h-RFeO$_3$ are trivalent and carry magnetic moments. 
 The magnetic moments on Fe come approximately from the electronic spins (for convenience, we use spin and magnetic moments interchangeably for the Fe sites to describe the magnetic order).  
 The strongest magnetic interaction between the Fe sites is expected to be the exchange interaction within the FeO layer, which can be written as:

\begin{equation}
H_{ex}^{ab}=\sum \limits_{i≠j,n} J_{i,j}^{ab} \vec{S}_i^{\frac{nc}{2}} \cdot \vec{S}_j^{\frac{nc}{2}},
\end{equation}

 where $\vec{S}_i^{n\frac{c}{2}}$ is the spin on the $i$th Fe site in the $z=\frac{nc}{2}$ FeO layer, $n$ is an integer, and $J_{i,j}^{ab}$ is the exchange interaction strength between the $i$th and $j$th sites in the same FeO layer. 
 Due to the two dimensional triangular lattice and antiferromagnetic interaction between Fe sites, this interaction is frustrated if the spins are along the $c$-axis.    
 On the other hand, if the spins are within the FeO plane, the frustration is lifted, generating the so-called the 120 degree orders, as shown in Fig. \ref{Fig_spinstructures}. 
 In the 120-degree antiferromagnetic orders, the moments in the same FeO layer can collectively rotate within the a-b plane, corresponding to the degree of freedom $\phi$ (Fig. \ref{Fig_spinstructures}).


As shown in Fig. \ref{Fig_RFeO3_structure}, there are two FeO layers in the unit cell of h-RFeO$_3$. The interlayer interactions determine the magnetic order along the $c$-axis, which can be written as:

\begin{equation}
H_{ex}^c=\sum \limits_{i,j,n} J_{i,j}^c \vec{S}_i^{\frac{nc}{2}}\cdot \vec{S}j^{\frac{(n+1)c}{2}},
\end{equation}

where $J_{i,j}^c$ is the exchange interaction strength between the $i$th and $j$th sites in the neighboring FeO plane. By combining the two independent directions of the spins in one FeO layer ($\phi$=0 or $\phi$=90 degrees) and the two different alignments between the spins in the two neighboring FeO layers (parallel or antiparallel), one can construct four independent magnetic orders ($\Gamma_1$ to $\Gamma_4$), as shown in Fig. \ref{Fig_spinstructures}. For h-RFeO$_3$, the spins follow one of the orders in $\Gamma_1$ to $\Gamma_4$ or a combination.

\subsubsection{Single ion anisotropy}
The key for the 120-degree magnetic orders is that the spins on the Fe sites lie in the a-b plane. This spin orientation is affected by the single ion anisotropy, which can be written as:

\begin{equation}
H_{SIA} =\sum \limits_i a_z (\vec{S}_i\cdot \vec{c})^2+(\vec{S}_i\cdot \hat{n})^2,
\end{equation}

 where the sign of $a_z$ determines whether the $c$-axis is an easy axis or a hard axis and $a_n$ and $\hat{n}$ ̂determine the preferred value of the angle $\phi$. 
 If $a_z$ has a large negative value, the electronic spins tend to point along the $c$-axis, corresponding to a magnetic frustration because of the antiferromagnetic interactions between the Fe sites; 
 which suppresses the magnetic order temperature. 
 In the limit of the Ising model (spins must be along $c$-axis), no long-range order can be formed. 
 Therefore, the single ion anisotropy is an important parameters that affects the magnetic ordering.


 If the magnetic order is $\Gamma_2$ or $\Gamma_3$, the Dzyaloshinskii-Moriya (DM)\cite{Dzyaloshinsky1958,Morin1950} interaction will cause a canting of the spins on the Fe sites toward the $c$-axis;
 this generates the non-zero components of the spins along the $c$-axis. 
 The DM interaction can be written as:

\begin{equation}
 H_{DM} = \sum \limits_{i,j,n}  \vec{D}_{i,j} \cdot ( \vec{S}_i^{\frac{nc}{2}} \times \vec{S}_j^{\frac{nc}{2}}),
\end{equation} 

 where $D_{i,j}$ is the vector interaction strength. 
 In particular, in the $\Gamma_2$ magnetic order, the $c$-axis component of the spins on all Fe sites are parallel; 
 this generates a net magnetic moment along the $c$-axis, causing a spontaneous magnetic polarization (weak ferromagnetism).

\section{Stabilization of RFeO$_3$ in hexagonal structure}
 Although Fe$^{3+}$ and Mn$^{3+}$ have almost identical radius, RFeO$_3$ is stable in orthorhombic structuer, while RMnO$_3$ is stable in hexagonal structure.
 According to the discussion in \ref{marker:stability}, the stability of the hexagonal manganites is a special case, while the stability orthorhombic structure for RFeO$_3$ follows closer to the trend of other rare earth transition metal oxides.
 Nevertheless, the metastable hexagonal structure in RFeO$_3$ can be achieved using special material preparation methods.
 These methods are: 
 1) wet-chemical method,\cite{Yamaguchi1991,Mizoguchi1996,Li2008,Hosokawa2009,Hosokawa2011,Downiea2012}
 2) under-cooling from a melt,\cite{Nagashio2002,Nagashio2006,Kumar2008,Kuribayashi2008,Magome2010,Kumar2013}
 3) thin film growth on substrates with trigonal symmetry,\cite{Bossak2004,Akbashev2011,Iida2012,Jeong2012,Pavlov2012,Akbashev2012b,Jeong2012,Jeong2012b,Wang2012,Wang2013,Roddatis2013,Moyer2014}
 and 
 4) doping either in the R site or in the Fe site.\cite{Magome2010,Masuno2013}
 In the method 1)-3), interface energy between the crystalline phase and the liquid (or amorphous phase) is the key; in the method 4), the reduction of the Gibbs free energy of the crystalline phase itself is employed.

 Although the orthorhombic structure is the ground state structure for RFeO$_3$, the hexagonal structure is an intermediate metastable state between the liquid (amorphous) and the crystalline orthorhombic state.
 Consider a transition from a liquid (amorphous) phase to a crystalline phase, the nucleation of the crystalline phase generates an interface between the two phases.
 The competition between the energy gain in forming the crystalline phase and the energy loss in forming the interface results in a critical size of the nucleation for the growth of the crystalline phase; this critical size corresponds to an energy barrier:

\begin{equation}
\Delta G^*=\frac{16 \pi \sigma^3 v_c^2}{3 \Delta \mu^2}, 
\label{Eq_barrier}
\end{equation}

 where $\sigma$ is the surface energy, $v_c$ is the molar volume, $\Delta\mu$ is the molar change of chemical potential.

 Because the orthorhombic phase is the stable crystalline phase, the $\Delta \mu^2$ is larger for the liquid$\rightarrow$ orthorhombic transition.
 On the other hand, if the interface energy $\sigma$ between the liquid and hexagonal phase is smaller, $\Delta G^*$ can be smaller for the liquid$\rightarrow$ hexagonal transition in a certain temperature range, considering that $\Delta\mu$ decrease as temperature increases. 
 In fact, the symmetry of hexagonal phase is higher than that of the orthorhombic phase, which suggests a smaller entropy change between the liquid and the hexagonal phase, or a smaller interface energy.
  A simulation on the energy shows that below a certain temperature, the energy barrier for forming hexagonal phase can be lower than that for forming orthorhombic phase.\cite{Nagashio2002}
 The temperature range becomes narrower for larger R$^{3+}$ and diminishes in LaFeO$_3$.

 This model has been corroborated by experimental observations.
 First, hexagonal ferrites are formed in an undercooled melt (from above 2000 K by laser-heating) of RFeO$_3$. 
 For YFeO$_3$, the hexagonal phase exists as a transient state, while for LuFeO$_3$, the hexagonal phase persist even down to the room temperature.\cite{Nagashio2002,Nagashio2006}
 Second, using wet-chemical method, amorphous YFeO$_3$, EuFeO$_3$ and YbFeO$_3$ are first created; 
 after annealing to above $750\,^{\circ}\mathrm{C}$, nanoparticles (10-50 nm diameter) of hexagonal phase were generated; 
 further annealing converts the hexagonal phase into the orthorhombic phase.\cite{Yamaguchi1991}

 Therefore, the smaller $\sigma$ at the liquid/hexagonal interface (compared to that at the liquid/orthorhombic interface) is the key in making the hexagonal phase metastable during the liquid$\rightarrow$crystalline solid transition. 
 On the other hand, as the size of the hexagonal crystalline phase grow (3 dimensionally), the interface energy becomes less important; 
 this is why the hexagonal phase exists in the form of nanoparticles or in the form of impurity phase (or even transient phase) in bulk samples.
 To maximize the effect of interface energy in stabilizing the hexagonal phase, epitaxial thin film growth employs the substrate surface of trigonal or hexagonal symmetry which enlarges the contrast between the $\sigma$ of the substrate/hexagonal and substrate/orthorhombic interfaces.
 
 In the epitaxial growth of thin films, the material grows along only one dimension (perpendicular to the surface), which makes the interface energy between the film material and the substrate much more important.
 Here the total Gibbs free energy of the substrate and the film is

\begin{eqnarray}
G=A (\sigma+gd), \\ \notag
\label{Eq_Gibbs_film}
\end{eqnarray}
 where $g=\frac{\mu}{v_c}$ is the Gibbs free energy per unit volume of the film, $A$ and $d$ are the surface area and film thickness.

 The difference of the Gibbs free energy between the substrate/hexagonal and substrate/orthorhombic combinations is

\begin{eqnarray}
\Delta G=A (\Delta \sigma+\Delta gd), \\ \notag
\end{eqnarray}
 Therefore, if the a substrate of trigonal or hexagonal symmetry is used to minimize the substrate/hexagonal interface energy $\sigma$, the combination of the hexagonal phase and the substrate actually can be the stable state. 
 In other words, below a critical thickness
\begin{equation}
d_c=-\frac{\Delta \sigma}{\Delta g},
\end{equation}
 the hexgonal phase is more energy favorable.
 For $d>d_c$, the hexagonal phase will remain metastable because the energy barrier for forming the hexagonal phase will always be lower than that for the forming the orthorhombic phase if the underlayer is hexagonal.
 The thickness of the h-RFeO$_3$ film in the literature is typically less than 100 nm on YSZ and Al$_2$O$_3$ substrates.
 Recently a growth of 200 nm h-LuFeO$_3$ on YSZ using oxide molecular-beam epitaxy is reported.\cite{Moyer2014} 
 However, a systematic study on the critical thickness to investigate the dependence on the substrate and the type of R$^{3+}$ is still needed.

\section{Thin film growth in hexagonal structure}
 The epitaxial growth of thin films depends greatly on the processes, which may generate the stable and metastable states of the substrate/film heterostructures.
 Here, we will not consider the growth dynamics and growth methods and ignore the morphology of the epitaxial thin films.
 Instead, we focus on the equilibrium state of the epitaxy, assuming the stoichiometry of the epilayer is correct.

\subsection{Epitaxial orientations}

 The epitaxial growth of a thin film on top of a substrate is characterized by the parallelism of the contact plane (texture orientation) and the parallelism of the crystallographic direction (azimuthal orientation). 
 The epitaxial orientation is determined by the minimum free energy conditions, which relate to the bonding across the substrate-epilayer interface and the lattice misfit.
 
 According to the Royer's rules\cite{Royer1928} for epitaxial growth of ionic crystals, the crystal planes in contact must have the same symmetry. 
 Therefore, if the surface symmetry of the substrate is triangular, the epilayer of RFeO$_3$ is either h-RFeO$_3$ (0001) or o-RFeO$_3$ (111) in the pseudo cubic coordinates, depending on the interfacial energy which is determined by the difference in structures and the lattice misfit.
 For example, if SrTiO$_3$ (111) is used as the substrate, the o-RFeO$_3$ will be the favorable structural of the epilayer because o-RFeO$_3$ share the similar (perovskite) structure and the similar lattice constants ($\sim$2\% misfit); this is actually a case close to homoepitaxy.\cite{Markov2003}
 On the other hand, the lack of substrate of similar structure to h-RFeO$_3$ makes the homoepitaxy very difficult.
 In most case of the epitaxial growth of h-RFeO$_3$ films, although the substrates have triangular symmetry, the structure and lattice constants are very different from h-RFeO$_3$, which certainly falls into the category of heteroepitaxy.\cite{Markov2003}

\begin{table*}[ht]
\addtolength{\tabcolsep}{-6pt}
\caption{
 The structures of the substrates and epitaxial orientations with h-RFeO$_3$.
 The lattice constants of the h-LuFeO$_3$ are a=b=5.9652 \AA$ $ and c=11.7022 \AA. \protect\cite{Magome2010}}
\label{tab:LFO_substrates}\centering

\begin{tabular}
{@{}ccccc@{}}\toprule
Substrate&Structure&Lattice Constants&Lattice Constant&Epitaxial Orientation \\
 & & (Bulk, in \AA )&(In-plane, in \AA)&(substrate $\parallel$ h-LuFeO$_3$)\\
\colrule
Al$_2$O$_3$(0001) & R$\bar{3}$c(167) &a=4.7602, c=12.9933 & a’=b’=4.7602 & $(0001)\langle 100\rangle ||(0001) \langle 100\rangle$\\
YSZ(111) & Fm$\bar{3}$m(225) & a=5.16(2) & a’=b’=7.30 & $(111) \langle $11-2$ \rangle ||(0001) \langle 100 \rangle$\\
Pt(111) & Fm$\bar{3}$m(225) & a=3.9231 & a’=b’=5.548 & $(0001) \langle $1-10$ \rangle ||(0001) \langle 100 \rangle$\\
\botrule
\end{tabular}
\end{table*}

\begin{figure}[ht]
\centerline{
\includegraphics[width = 4 in]{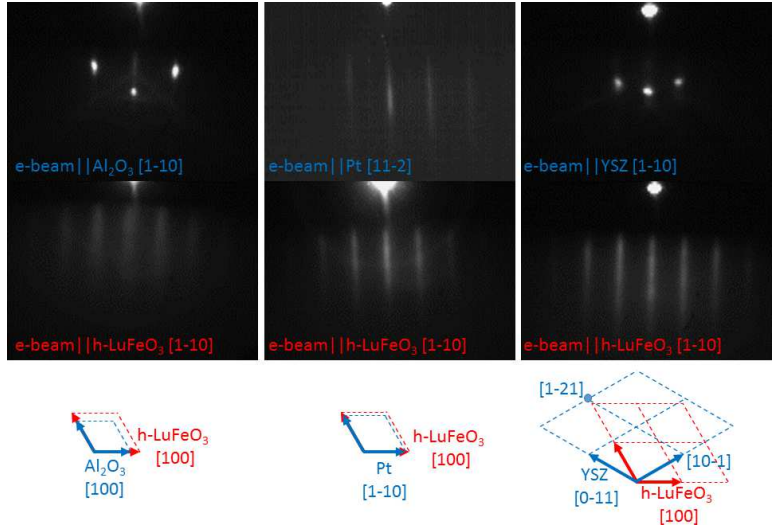}
}
\caption{ (Color online)
	RHEED Pattern of the substrates and the overlayer of h-LuFeO$_3$ .
}
\label{Fig_rheed-3substrates}
\end{figure}

\begin{figure}[ht]
\centerline{
\includegraphics[width = 2.5 in]{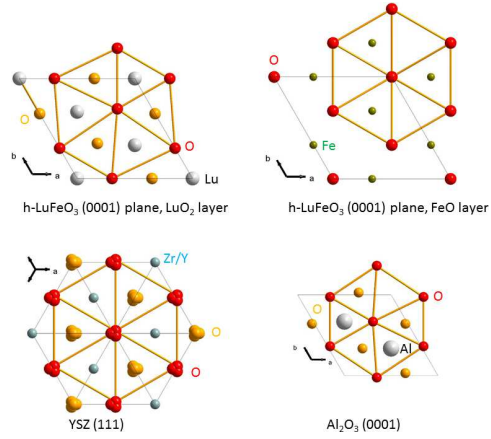}
}
\caption{ (Color online)
	The possible interfacial layers of the substrates and h-RFeO$_3$.
	The red (orange) color of the oxygen indicates that the atoms are above (below) the center of mass along the surface normal.
	The lattices of h-RFeO$_3$, Al$_2$O$_3$, and YSZ are in the same scale.
	The multiple oxygen atoms on the same site indicates the uncertainty of the oxygen positions in YSZ. 
	The projection of the edges of the bulk unit cells of the structures are indicated by the thin lines.
	The oxygen networks are highlighted by the connection between the oxygen atoms with the yellow rods.
}
\label{Fig_interfaces}
\end{figure}

 Epitaxial h-RFeO$_3$ thin films have been deposited on various substrates including Al$_2$O$_3$ (0001), yttrium stabilized zirconium oxide (YSZ) (111) and Al$_2$O$_3$ (0001) buffered with Pt (111) layers.\cite{Bossak2004,Akbashev2011,Iida2012,Jeong2012,Pavlov2012,Akbashev2012b,Jeong2012,Jeong2012b,Wang2012,Wang2013,Roddatis2013,Moyer2014}
 All the epitaxial growth is along the $c$-axis of h-RFeO$_3$, to satisfy the matching triangular symmetry.

 As shown in Table \ref{tab:LFO_substrates}, there is no obvious lattice match between the h-RFeO$_3$ and Al$_2$O$_3$, YSZ, or Pt. 
 On the other hand, the epitaxial growth can be obtained, as shown in Fig. \ref{Fig_rheed-3substrates} as an example using the pulsed laser deposition.
 The in-plane epitaxial orientations is normally studied using x-ray diffraction.
 Nevertheless, the reflection high energy electron diffraction (RHEED) provides an in-situ, time-resolved characterization of the in-plane epitaxial orientation too.
 As shown in Fig. \ref{Fig_rheed-3substrates}, one can calculate the in-plane lattice constants from the separation of the RHEED patterns and determine the orientation of the overlayer with respect to the substrates.
 The resulting epitaxial orientations are shown in Table \ref{tab:LFO_substrates}.

 It appears that the key in growing epitaxial h-RFeO$_3$ films is the substrate with a surface of triangular symmetry.
 On the other hand, the in-plane 2-dimensional lattice constants between h-RFeO$_3$ and the substrates do not show obvious match.
 For Al$_2$O$_3$, the in-plane 2-dimensional lattice constants are a=b=4.7602 \AA, which has a lattice misfit greater than 25\%.
 If one considers the super cell match (four times of lattice constant of h-LuFeO$_3$ v.s. five times of lattice constant of Al$_2$O$_3$), the lattice misfit is -0.25\%, which falls into the category of coincident lattices.\cite{Markov2003}
 There is also a large (-18\%) lattice misfit between h-LuFeO$_3$ (0001) and YSZ (111), if we take YSZ [1-10] as the direction of in-plane 2-dimensional basis.
 Using a super-cell scheme (five times of lattice constant of h-LuFeO$_3$ v.s. four times of lattice constant of YSZ), the lattice misfit is 2\%.
 It turns out that the in-plane epitaxial orientation between the h-LuFeO$_3$ and YSZ substrates is $\langle 100 \rangle$ of h-LuFeO$_3$ $||$ $\langle11-2\rangle$ of YSZ, which does not follow the super-cell scheme, but involves a 30 degree rotation from that, which is similar to the observed epitaxy of YMnO$_3$ (0001) on YSZ (111).\cite{Dubourdieu2007,Wu2011}
 After the 30 degree rotation, the lattice point of [1-21] of YSZ is close to the lattice point of [020] of h-LuFeO$_3$, with a misfit of -5.6\%, as shown in Fig. \ref{Fig_rheed-3substrates}.
 For the growth of h-LuFeO$_3$ on Pt (111), the lattice misfit between h-LuFeO$_3$ (0001) and Pt (111) is -7.5\%, if we take Pt [1-10] as the direction of in-plane 2-dimensional basis.
 The growth turns out to have the epitaxial orientation of  $\langle 100 \rangle$ of h-LuFeO$_3$ $\parallel$ $\langle11-2\rangle$ of Pt, despite the large misfit, which is understandable because the 30 degree rotation will generate only a much larger misfit.

 Since the azimuthal epitaxial orientations of h-LuFeO$_3$ films cannot be explained simply by the lattice misfit, the detailed interfacial structure must be responsible.
 As shown in Fig. \ref{Fig_interfaces} (also in Fig. \ref{Fig_RFeO3_structure}), there are two types of layers in h-RFeO$_3$: the LuO$_2$ layer and the FeO layer. 
 The common part of the LuO$_2$ and FeO layers are the plane of oxygen network with triangluar connectivity. 
 The YSZ and Al$_2$O$_3$ structures also contain plane of oxygen triangular lattice.
 It appears that the epitaxial orientations of the h-LuFeO$_3$ films on YSZ (111) and Al$_2$O$_3$ (0001) aligns the oxygen networks.
 In particular, the rotation of YSZ in-plane axis by 30 degree is explained: after the rotation, the oxygen networks of h-LuFeO$_3$ (0001) and YSZ (111) may match at the interface with a -5.6\% misfit.
 The share of the oxygen network at the interface reduces the total energy, which may be responsible for the observed azimuthal epitaxial orientation.
 The better match of the oxygen networks between the h-LuFeO$_3$ (0001) and YSZ (111) in comparison with Al$_2$O$_3$ (0001) also indicates that the interfacial bonding of h-LuFeO$_3$(0001)/YSZ(111) is stronger.
 If the key in the epitaxy is the share of oxygen network, one can infer that it is more energetically favorable to have the LuO$_2$ layer (instead of the FeO layer) at the interface with the YSZ and Al$_2$O$_3$ substrates, which means that the surface of the h-LuFeO$_3$ films will have FeO layer as the termination.

\subsection{Strain at the interface}

\begin{figure*}[ht]
\centerline{
\includegraphics[width = 5 in]{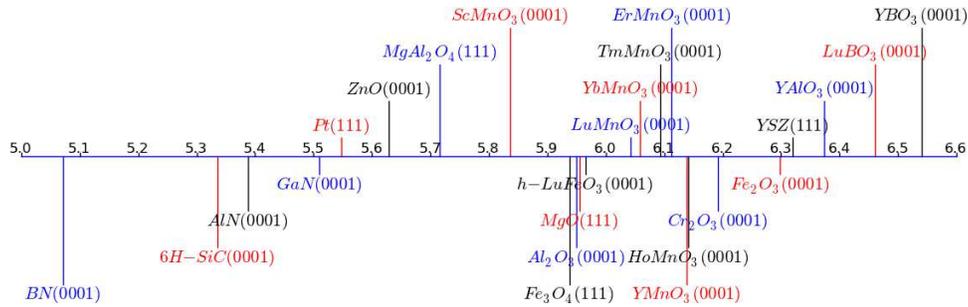}
}
\caption{ (Color online)
 A spectrum of the effective in-plane lattice constants (in \AA) of different surfaces. 
 For the corundum materials (Al$_2$O$_3$, Fe$_2$O$_3$, and Cr$_2$O$_3$), the values are 5/4 of the original in-plane lattice constant; 
 for spinel (MgAl$_2$O$_4$ and Fe$_3$O$_4$), the values are 1/2 of the in-plane lattice constant; 
 for the hexagonal material (6H-SiC, GaN, AlN, BN, ZnO, YAlO$_3$, LuBO$_3$, and YBO$_3$), the values are the in-plane lattice constant with a 30-degree rotation; 
 for YSZ, the values are 1/2 the in-plane lattice constant after a 30-degree rotation. 
}
\label{Fig_interfaces}
\end{figure*}

 Epitaxial strain is an extremely important issue in epitaxial thin film growth, because the strain may change the properties of the epilayer, offering opportunities of material engineering.
 Here we discuss the possibility of straining the h-RFeO$_3$ film, although no systematic work on the strain effect has been reported on these films.

 Besides the epitaxial orientations discussed above, the interfacial structure is another key property of the epitaxy.
 Depending on the relative strength between the interfacial bonding ($\psi_{s-e}$) and the bonding in the epilayer ($\psi_e$) and the lattice misfit, the interfacial structure can be divided as the following categories, assuming the substrate is rigid (bonding in the substrates $\psi_s=\infty$).
 1) When the $\psi_{s-e}$ is much weaker than $\psi_e$, the interface is a vernier of misfit.\cite{Markov2003} 
 The substrate and the epilayer maintain their structures; there is minimal strain on the epilayer.
 2) When the $\psi_{s-e}$ is much stronger than $\psi_e$, the interface has a homogeneous strain.
 The epilayer follows the structural of the substrates; the epitaxy is pseudomorphic.
 3) When the $\psi_{s-e}$ is comparable to $\psi_e$, the misfit is accommodated by a mixture of locally homogeneous strain and dislocations.
 Over all, the epitaxy is called misfit dislocation with periodic strain.
 Ideally, the category 2), the homogeneous strain is desirable for material engineering.
 However, the category 3), misfit dislocations, corresponding to a gradient of strain along the out-of-plane direction, also causes a change of material properties.
 On average, the lattice parameter of the epitaxial film with misfit dislocations will be in between the value of the substrate and the bulk value of the film material, which can be consider as partially strained.

 A qualitative criteria for the homogeneous stain can be discussed using a one-dimensional model, in which the substrate is parameterized using a sinusoidal potential (period $a_s$, amplitude $W$), and the epilayer is parameterized as a chain of particles connected by springs of natural length $a_e$ and spring constant $k$;
 the misfit is defined as $f=\frac{a_e-a_s}{a_s}$.\cite{Markov2003}
 The maximum misfit allowed for a homogeneous strain was found as $f_s=\frac{2}{\pi\lambda}$, where $\lambda=\sqrt{\frac{k a_e^2}{2W}}$.
 Assuming that multiple epilayers can be described simply by replacing $k$ with $nk$, where $n$ is the number of epilayers, one can find the maximum number of epilayers that allows for homoegeneous strain as $n=(\frac{2}{f\pi\lambda})^2$.
 The bottom line of this model is that the homogeneous stain occurs only when the misfit is small and the interfacial bonding is strong.

 Based on this qualitative model, one can discuss the possible homogeneous strain in h-RFeO$_3$ films.
 For the Al$_2$O$_3$ (0001) substrates, the lattice mismatch of the super cell is small;
 but the huge misfit of the lattice constant suggests weak interfacial bonding.
 For YSZ (111) substrate, the similar oxygen network suggests strong interfacial bonding.
 But the large misfit ($\sim$5\%) makes the homogeneous strain more difficult.
 Therefore, the homogenous strain in h-RFeO$_3$ on these substrates will have a very small critical thickness.

 Partially strained films have been reported for epitaxial growth of YMnO$_3$ (0001) on YSZ (111).\cite{Wu2011}
 In contrast, the strain effect of h-LuFeO$_3$ (0001) on YSZ (111) is minimal.
 A possible model in terms of the difference in the electronic structure of RMnO$_3$ and h-RFeO$_3$ has been proposed to account for the observation.\cite{Bossak2004}
 In this model, high flexibility of the RMnO$_3$ lattice is hypothesized based on the 3d$^4$ configuration of Mn$^{3+}$. 
 In contrast, the 3d$^5$ configuration of Fe$^{3+}$ makes the h-RFeO$_3$ lattice more rigid.
 So the $\psi_e$ and the effective spring constant $k$ in h-RFeO$_3$ are large, making these materials more difficult to strain and the critical thickness is small.
 Note also that the lattice misfit between RMnO$_3$ (0001) and YSZ (111) is about half of that between h-RFeO$_3$ (0001) and YSZ (111), which may also be an important factor of increased critical thickness in RMnO$_3$/YSZ, according to the model discussed above.

 Fig. \ref{Fig_interfaces} shows a spectrum of the effective in-plane lattice constants of surfaces of triangular symmetry, calculated from the corresponding bulk structure.
 Epitaxial growth of h-RFeO$_3$ on most of the materials as substrates have not been reported, which can be explained by the lack of good lattice match.
 If a substrate of RMnO$_3$ is used, the homogeneous strain may be possible because the lattice mismatch is small and the interfacial bonding is strong due to the same P6$_3$cm lattice structure.
 
\section{Ferroelectricity}

\begin{figure*}[ht]
\centerline{
\includegraphics[width = 3.5 in]{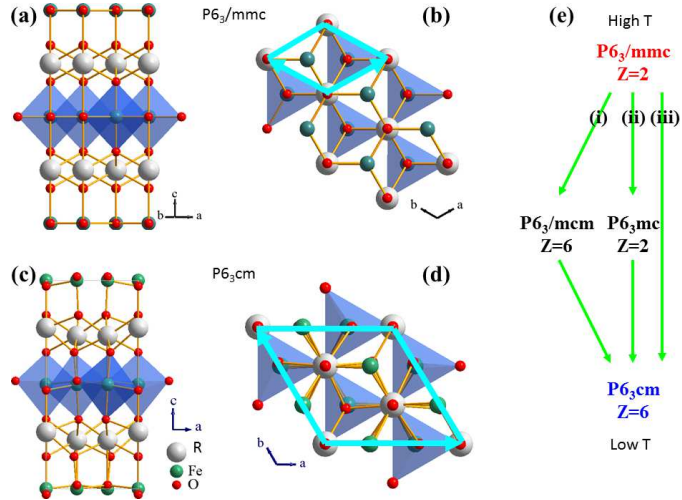}
}
\caption{ (Color online)
 Structure of the h-RFeO$_3$ at room temperature (a), (b) with P6$_3$cm symmetry and at high temperature (c), (d) with P6$_3$/mmc symmetry. 
 (e) The possible routes of structure transitions.
 After Wang et. al. 2013.\protect\cite{Wang2013}
}
\label{fig_FE_structure}
\end{figure*}

\begin{figure*}[ht]
\centerline{
\includegraphics[width = 3.5 in]{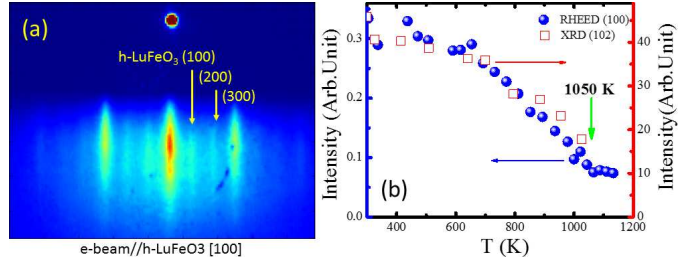}
}
\caption{ (Color online)
 Structural characterization of the h-LuFeO$_3$ films.
 (a) A RHEED image of the h-LuFeO$_3$ film at 300 K with an electron beam along the h-LuFeO$_3$ [100] direction. 
 (b) Intensities of the RHEED (100) peak and XRD (102) peak as functions of temperature. 
 After Wang et. al. 2013.\protect\cite{Wang2013}
}
\label{fig_FE_RHEED}
\end{figure*}

\begin{figure*}[ht]
\centerline{
\includegraphics[width = 3.5 in]{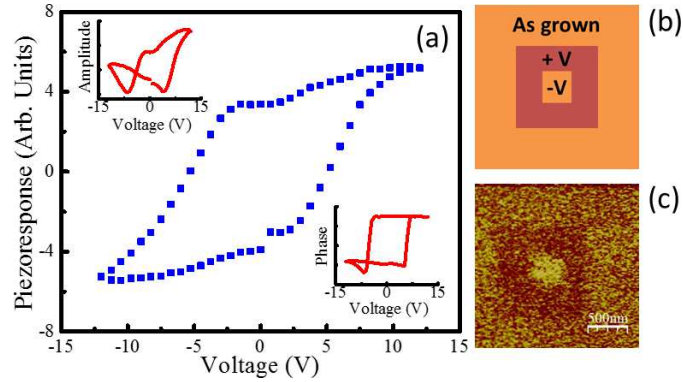}
}
\caption{ (Color online)
  Piezoelectric polarization of h-LuFeO$_3$ films.
  (a) PFM response displaying square-shaped hysteresis loop. 
  The amplitude and phase are shown in the insets.
  (b) Schematic of written domain pattern with DC voltage ({\it V} = 20 V$_{dc}$).
  (c) PFM image of the same region without DC voltage.
	After Wang et. al. 2013.\protect\cite{Wang2013}
}
\label{fig_FE_PFM}
\end{figure*}

 The transition from the high temperature paraelectric phase to the ferroelectricity phase in h-RFeO$_3$ involves change of structure, including the symmetry and size of the unit cell.
 As shown in Fig. \ref{fig_FE_structure}, from the high temperature P6$_3$/mmc structure to the low temperature P6$_3$cm structure, there are three possible routes, involving different intermediate structures.
 In route (i), K$_1$ mode freezes first, followed by the freezing of $\Gamma_2^-$ and K$_3$; 
 this generates an intermediate non-polar structure with trippled unit cell.
 In route (ii), $\Gamma_2^-$ mode freezes first, followed by the freezing of K$_3$;
 this generates an intermediate ferroelectric structure with the same unit cell size as the high temperature structure.
 In route (iii), $\Gamma_2^-$ and K$_3$ freeze simultaneously, with no intermediate state.

 The structural transition of h-LuFeO$_3$ has been studied up to 1150 K using x-ray and electron diffraction.\cite{Wang2013}
 As shown in Fig. \ref{fig_FE_RHEED}, the room temperature RHEED pattern shows intense diffraction streaks separated by weak streaks, which can be understood in terms of the detailed structure of h-LuFeO$_3$.
 The freezing of K$_3$ mode corresponds to a $\sqrt{3}\times\sqrt{3}$ reconstruction in the a-b plane (tripling the unit cell).
 The RHEED pattern is in perfect agreement with the $\sqrt{3}\times\sqrt{3}$ reconstruction in the a-b plane. 
 From the streak separation, the in-plane lattice constants of the h-LuFeO$_3$ films were estimated as to be consistent with the value of the bulk P6$_3$cm structure.
 Hence, the indices of the diffraction streaks can be assigned using a P6$_3$cm unit cell, as indicated in Fig. \ref{fig_FE_RHEED}(a).
 By measuring the diffraction intensities of the (102) peak of XRD and (100) peak of RHEED as a function of temperature, the structural transition at which the trippling of unit cell disappears, was determined to occur at 1050 K.
 The piezoelectric reponse has been demonstrated in h-LuFeO$_3$ films, as shown in Fig. \ref{fig_FE_PFM}, using piezoforce microscopy (PFM).\cite{Wang2013}
 Switching of the polarization direction at room temperature is also achieved using a conducting tip on a h-LuFeO$_3$ film grown on Al$_2$O$_3$ (0001) buffer with Pt.
 Combining the PFM study and the structural characterization, one may conclude that 
 1) the h-LuFeO$_3$ films are ferroelectricity at room temperature; 
 2) the polar structure and the trippling of unit cell persist at least up to 1050 K.

 Another evidence of the ferroelectricity of the h-LuFeO$_3$ films grown on Pt-buffered Al$_2$O$_3$ (0001) substrate, are in the temperature dependence of the electric polarization, measured between 300 and 650 K.\cite{Jeong2012b}
 A clear transition was observed at 560 K, which was attributed as the Curie temperature of the ferroelectricity, because a dielectric anomaly was also observed at the same temperature.
 The electric polarization at 300 K was determined as 6.5 $\mu$C/cm$^2$, which is comparable to that of YMnO$_3$.

 These experimental observations are consistent with the theoretical calculations.
 It has been shown by the density functional calculations that the origin of ferroelectricity in h-LuFeO$_3$ is similar to that in YMnO$_3$.\cite{Das2014}
 In other words, the instability of the $\Gamma_2^-$ mode is induced by the frozen K$_3$ mode. 
 So h-LuFeO$_3$ is intrinsically antidistortive, extrinsically ferroelectric (improperly ferroelectric).

\begin{figure*}[ht]
\centerline{
\includegraphics[width = 4 in]{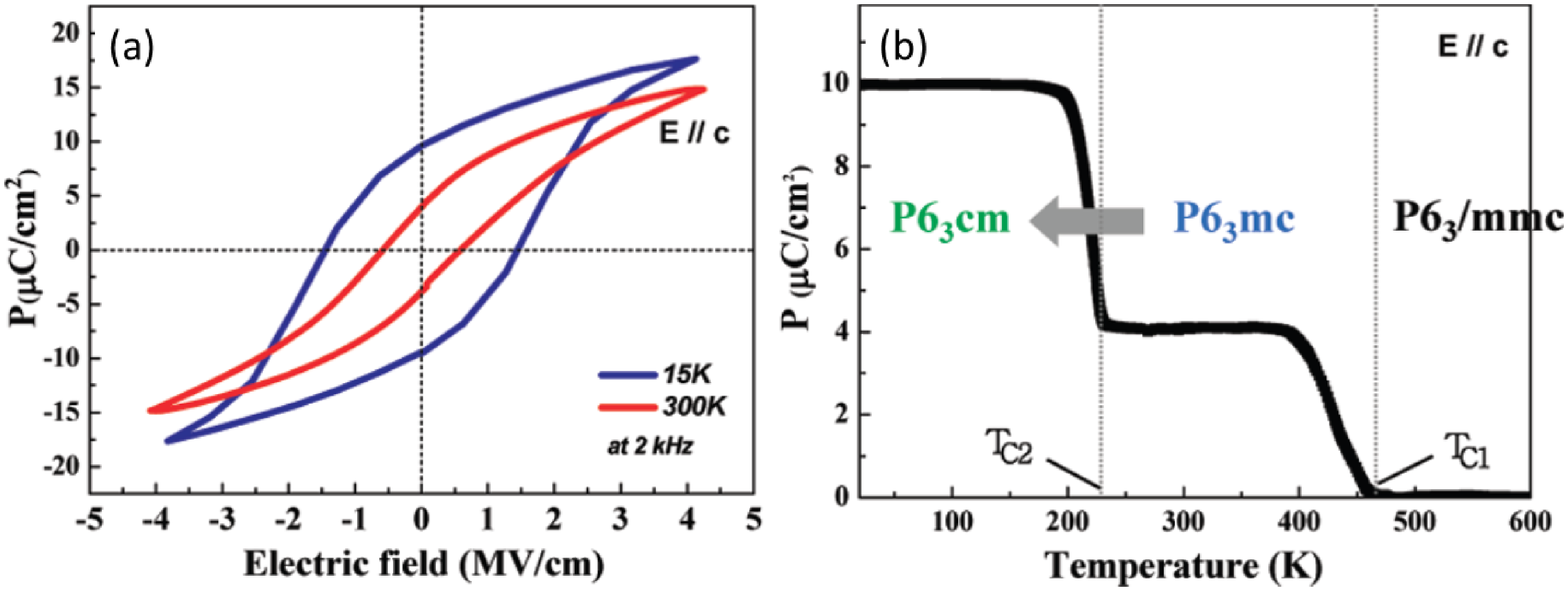}
}
\caption{ (Color online)
 Charcterization of the ferroelectricity in h-YbFeO$_3$ films.
 a) Polarization-electric field loop measured at two different temperatures.
 b) Temperature dependence of the electric polarization.
 After Jeong et. al. 2012.\protect\cite{Jeong2012}
}
\label{fig_FE_PEloop}
\end{figure*}

 The ferroelectric properties in h-YbFeO$_3$ have been studied on films grown on Al$_2$O$_3$ (0001) buffered with Pt by measuring the electric polarization.\cite{Jeong2012}
 The polarization-electric field loop was demonstrated to verify the ferroelectricity of h-YbFeO$_3$ (Fig. \ref{fig_FE_PEloop}(a). 
 The most interesting finding in h-YbFeO$_3$ is the two-step transition in the temperature dependence of the electric polarization.
 As shown in Fig. \ref{fig_FE_PEloop} (b), the ferroelectric polarization becomes non-zero below 470 K and a second transition occurs at 225K; 
 the low-temperature electric polarization was determined to be 10 $\mu$C/cm$^2$. 
 A two-step structural transition sequence is proposed as P6$_3$/mmc $\rightarrow$ P6$_3$mc $\rightarrow$ P6$_3$cm, based on the two observed transitions in the electric polarizations.
 This sequence corresponds to route ii) in Fig. \ref{fig_FE_structure}.
 The most intriguing inference here is that h-YbFeO$_3$ is properly ferroelectric with a P63mc structure at room temperature, which is difference from that of RMnO$_3$.
 On the other hand, reconstruction-fashioned 3-fold periodicity was observed in electron diffraction pattern at room temperature, indicating a trippling of the unit cell (compared with that of P6$_3$/mmc) in h-YbFeO$_3$, which contradicts the proposed P6$_3$mc structure at room temperature.\cite{Iida2012}
 Direct observation of the structural transition has not been demonstrated to occur at the transition of electric polarization.
 Other indication of the polar structure at room temperature is indicated by the optical second harmonic generation (SHG).\cite{Iida2012}
 In particular, a transition in the SHG signal was observed at 350 K.
 Dielectric anomaly was also observed in h-YbFeO$_3$ at this temperature, indicating ferroelectricity. 
   
 Although it is a consensus that the ferroelectric transition in h-RFeO$_3$ is accompanied by a structural transition, there is significant controversy in the literature about the symmetry of the structures and the transition temperature, which may be due to the sample-sample variation.
 A study on the structural and ferroelectricy transition of the same samples should be helpful in clarifying this issue.

\section{Magnetism}

 As discussed in Section \ref{marker:ferroelectricity}, the low temperature structure of h-RFeO$_3$ has the same space group P6$_3$cm as that of RMnO$_3$.
 This plus the assumption that the magnetic unit cell is the same as the structural unit cell, result in the four possible magnetic structures shown in Fig. \ref{Fig_spinstructures}, same as those in RMnO$_3$.
 The distinction of h-RFeO$_3$ compared with RMnO$_3$ is following:
 1) The spins of Fe$^{3+}$ are canted slightly out of the FeO plan, causing a weak ferromagnetism in the $\Gamma_2$ spin structure; this is reported for all the h-RFeO$_3$ regardless of the R$^{3+}$ site.\cite{Akbashev2011,Iida2012,Jeong2012,Jeong2012b,Jeong2012b,Wang2013,Moyer2014}
 2) When the R$^{3+}$ sites are magnetic, the moment of R$^{3+}$ will be aligned by the field of Fe$^{3+}$ sites, generating a large magnetic moment at low temperature; 
 the moment per formular unit (f.u.) is close to that of the R$^{3+}$.\cite{Iida2012,Jeong2012}
 3) The higher spin of Fe$^{3+}$ and the stronger exchange interactions between the Fe$^{3+}$ sites (compared with those of the Mn$^{3+}$) suggests higher magnetic ordering temperature $T_{\rm N}$.\cite{Das2014}


\begin{figure*}[ht]
\centerline{
\includegraphics[width = 2.5 in]{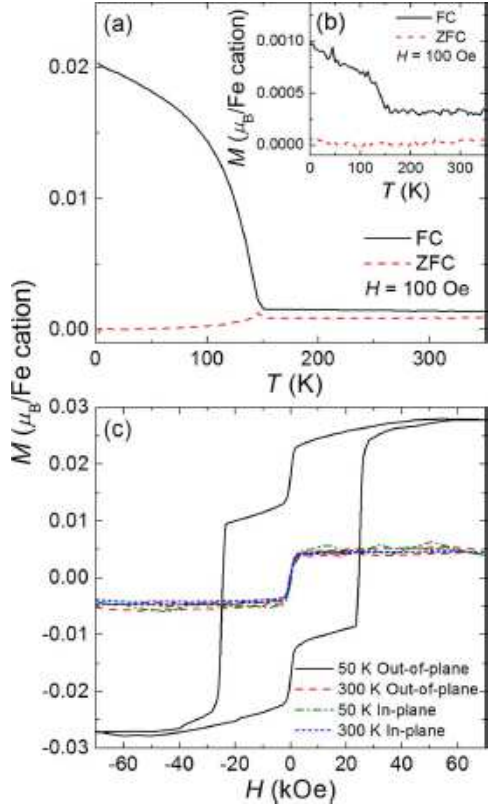}
}
\caption{ (Color online)
 Weak ferromagnetism of a nominally stoichiometric h-LuFeO$_3$ film.
 (a) Temperature dependence ($M-T$) of the magnetization in both field cool (FC) and zero field cool (ZFC) processes;
 the magnetic field is along the $c$-axis.
 (b) $M-T$ in both FC ZFC; the magnetic field is perpendicular to the $c$-axis.
 (c) The field depdendence of ($M-H$) when the magnetic field at different magnetic field directions and temperatures.
 After Moyer et. al. 2014.\protect\cite{Moyer2014}
}
\label{fig_mag_LFO-MH}
\end{figure*}

 The weak ferromagnetism in h-LuFeO$_3$ was first reported in films grown using metal-organnic chemical vapor deposition (MO-CVD) and confirmed later on in films grown using pulsed laser deposition and molecular beam epitaxy.\cite{Akbashev2011,Jeong2012b,Wang2013,Moyer2014}
 The origin of the weak ferromagnetism is attributed to the canting of magnetic moment of Fe$^{3+}$ because Lu$^{3+}$ has no magnetic moment.
 An important issue of the magnetism of h-LuFeO$_3$ films is the stoichiometry dependence.
 A non-stoichiometric h-LuFeO$_3$ film may contain LuFe$_2$O$_4$, Fe$_3$O$_4$ and Lu$_2$O$_3$ phases.
 These impurity phasese, particularlly LuFe$_2$O$_4$ and Fe$_3$O$_4$ may coexist epitaxially with the h-LuFeO$_3$ structure, which can introduce confusion in the magnetic characterization since both are ferrimagnetic (the Neel temperature are 240 K and 860 K for LuFe$_2$O$_4$ and Fe$_3$O$_4$ respectively).
 Indication of magnetic impurities was observed in Fe-rich h-LuFeO$_3$ films as the two-step transitions in the temperature dependence of the magnetization.\cite{Moyer2014}
 The transition at lower temperature is expected to be from h-LuFeO$_3$ because the higher transition is consistent with the formation of ferrimagnetism in h-LuFe$_2$O$_4$.
 An important observation is that the weak ferromagnetism of h-LuFeO$_3$ occurs at higher temperature when the Fe concentration is higher and saturate when the Fe and Lu ratio are close to one.

 Figure \ref{fig_mag_LFO-MH} shows the temperature and magnetic field dependence of the magnetization of a nominally stoichiometric h-LuFeO$_3$ film.\cite{Moyer2014}
 From the $M-T$ relation, the critical temperature for the weak ferromagnetism ($T_{\rm W}$) is determined as approximately 147 K, and the out-of-plane component of the Fe$^{3+}$ moment is 0.018 $\mu_B$/formula unit.
 From the low temperature $M-H$ relations, the coercive field of h-LuFeO$_3$ is found as approximately 25 kOe at 50 K.
 Therefore, there is a huge anisotropy in the magnetic moment of h-LuFeO$_3$; the $c$-axis is the easy axis.
 The sharp change of magnetization at 25 kOe in Fig. \ref{fig_mag_LFO-MH} indicate that the magnetic moments in h-LuFeO$_3$ are Ising-like, which is consistent the canting model for the origin of the weak ferromagnetism.
 Even in the nominally stoichiometric h-LuFeO$_3$ films, there appears to be a second magnetic component that persist beyond 350 K; this component also show anisotropy according to the $M-T$ relations.
 The true nature of this second component is however difficult to determine from the magnetometry in Fig. \ref{fig_mag_LFO-MH} because of the limited temperature range of the measurement.

\begin{figure*}[ht]
\centerline{
\includegraphics[width = 4 in]{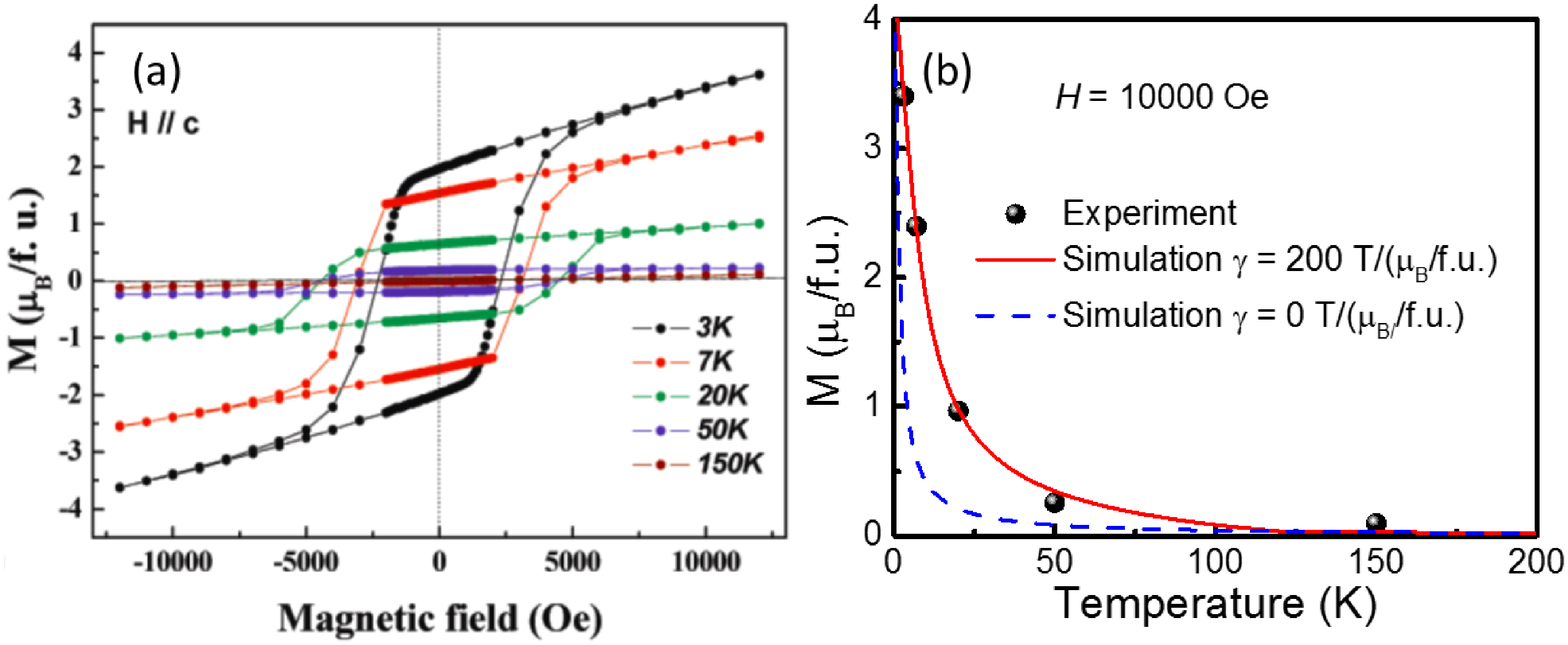}
}
\caption{ (Color online)
 Ferrimagnetism in h-YbFeO$_3$ films.
 (a) Magnetic-field dependence of the magnetization at various temperatures (after Jeong et. al. 2012).\protect\cite{Jeong2012}
 (b) Temperature dependence of the high-field magnetization.
 The dots are the experimental magnetization values taken from (a) at 10000 Oe.
 The lines are simulations.
}
\label{fig_Mag_YFO_MH}
\end{figure*}

 When the R$^{3+}$ site are magnetic (R=Dy-Yb), the total magnetic moment of h-RFeO$_3$ have contributions from both Fe$^{3+}$ and R$^{3+}$ sites.
 It turns out that the total moment of h-YbFeO$_3$ along the $c$-axis can be much larger than that of h-LuFeO$_3$.\cite{Iida2012,Jeong2012}
 As shown in Fig. \ref{fig_Mag_YFO_MH}(a), the magnetization of h-YbFeO$_3$ at 3 K in 10000 Oe  is close to 4 $\mu_B$ per formular unit, the magnetic moment of free Yb$^{3+}$.
 Therefore, the measured magnetization in Fig. \ref{fig_Mag_YFO_MH}(a) should come mainly from the moment of Yb$^{3+}$.
 According to the temperature dependence of the high-field magnetization [see Fig. \ref{fig_Mag_YFO_MH}(b)], the Yb$^{3+}$ sites show paramagnetic-like behavior.
 On the other hand, the onset of the field-cool magnetization suggests a correlation between the magnetization of the Yb$^{3+}$ sites with that of the Fe$^{3+}$ sites.
 Below, we model the effect of the magnetization of the Fe$^{3+}$ sites on that of the Yb$^{3+}$ sites, assuming that the interaction between the Yb$^{3+}$ sites are weak enough to be ignored.

 The Yb$^{3+}$ can be polarized by an effective magnetic field $B_{eff}$, which is a combination of the external field $H$ and the molecular field from Fe$^{3+}$ sites:
\begin{equation}
B_{eff}=\mu_0 H+\gamma M_{Fe},
\end{equation}
where $M_{Fe}$ is the magnetization of the Fe$^{3+}$ sites and $\gamma$ represents the strength of the interaction between Yb$^{3+}$ and Fe$^{3+}$ sites.
 Since the Yb$^{3+}$ sites are assumed to be isolated, the average projection of the magnetic moment along the $c$-axis, $\langle \mu_{Yb} \rangle $, follows the paramagnetic behavior described by the Langevin function: 
\begin{equation}\label{eq:Eq_YFO_Ferri}
\langle \mu_{Yb} \rangle = \mu_{Yb} L(\frac{\mu_{Yb} B_{eff}}{k_B T}), \\ 
\end{equation}
 where $L$ is the Langevin function, $\mu_{Yb}$ is the magnetic moment of Yb$^{3+}$, $k_B$ is the Boltzmann constant, and $T$ is temperature.
 Note that $M_{Fe}$ is temperature dependent;
 here we assume $M_{Fe}=M_{Fe}^0 {\rm cos}(\frac{\pi}{2} \frac{T}{T_W})$ for $T<T_W$, and $M_{Fe}=0$ for $T>T_W$.

 By fitting the experimental data in Fig. \ref{fig_Mag_YFO_MH}(b) at $H=10000 \: {\rm Oe}$ using Eq. (\ref{eq:Eq_YFO_Ferri}) with parameters $T_W=120$ K and $M_{Fe}^{0}=0.02 \: \mu_B$/f.u, we found $\mu_{Yb}^0=4.3 \: \mu_B$ and $\gamma=200 \: {\rm T}/ (\mu_B/f.u.)$;
 this corresponds to a molecular field of 4 T for $M_{Fe}$=0.02 $\mu_B$/f.u.

 The simulated magnetization (same as $\langle \mu_{Yb} \rangle$ in $\mu_B$/f.u.) at 10000 Oe is displayed in Fig. \ref{fig_Mag_YFO_MH}, using the parameters found from the fitting.
 A simulation assuming no molecular field from Fe$^{3+}$ site ($\gamma$=0) is also shown for comparison.
 It is clear that the moments of Yb$^{3+}$ sites are significantly polarized by the molecular field of Fe$^{3+}$.
 In other words, the weak spontaneous magnetization from the Fe$^{3+}$ sites are magnified by the existence of large, isolated Yb$^{3+}$ moments, which generates a huge residual magnetization at low temperature, as displayed in Fig. \ref{fig_Mag_YFO_MH}(a).

\begin{figure*}[ht]
\centerline{
\includegraphics[width = 4 in]{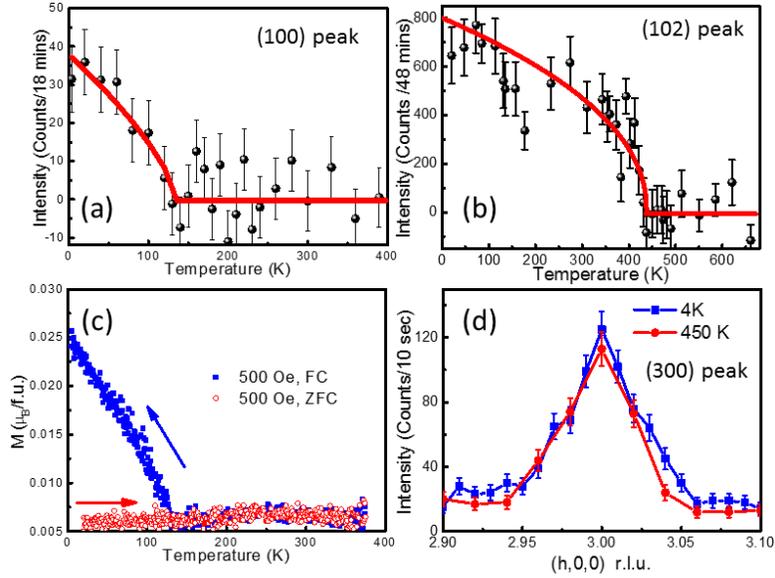}
}
\caption{ (Color online)
 Indication of high-temperature antiferromagnetism in h-LuFeO$_3$ films.
 Temperature dependence of the intensities of the (100) (a) and (102) (b) peaks of the neutron diffractions. 
 (c) The FC and ZFC magnetization along the $c$-axis as a function of temperature.
 (d) Scan of diffraction peak (300) in reciprocal lattice unit (r.l.u.) at 4 and 450 K.
 After Wang et. al. 2013.\protect\cite{Wang2013}
}
\label{fig_mag_LFO_Neutron}
\end{figure*}

 As discussed in Section \ref{marker:intromagnetism}, the magnetic orders in h-RFeO$_3$ are essentially 120-degree antiferromagnetic.
 Among the four possible magnetic structures, the DM interaction coefficient $\vec{D}_{i,j}$ is non-zero only for $\Gamma_2$ and $\Gamma_3$.
 The alignment between neighboring FeO layers are ferromagnetic and antiferomagnetic in $\Gamma_2$ and $\Gamma_3$ respectively.
 Therefore, the $\Gamma_2$ structure can be determined using magnetometry because of the non-zero spontaneous magnetization, which corresponds to the weak (parasitic) ferromagnetism discussed above.
 On the other hand, the other antiferromagnetic structures are more difficult to identify using magnetoemtry alone, particularly in thin film samples which contain small amount of materials.
  
 Neutron diffraction offers a way of measuring the magnetic orders without the need for a net magnetization, which is more suitable for antiferromagnetic materials.\cite{Lovesey1984}
 In addition, different selection rules of different diffraction peaks may provide more information on the magnetic structures. 
 
 Because the crystallographic and magnetic structure of h-RFeO$_3$ have the same unit cell, in general, the intensities of the neutron diffraction peaks have contributions from both nuclear and magnetic interactions between the atoms and the neutron beams. 
 However, the intensity could be dominated by nuclear or magnetic diffraction because of the difference between the two kinds of interactions: 
 1) all the sites contribute to the nuclear diffraction while only magnetic sites (Fe$^{3+}$ in the case of h-LuFeO$_3$) contribute to the magnetic diffraction; 
 2) the selection rules are different for the two interactions.
 One way to identify the contribution of the intensities is to examine the temperature dependence of the peak intensities and compare with that of the x-ray and electron diffraction which are not expected to reflect the magnetic interactions.
 As shown in Fig. \ref{fig_mag_LFO_Neutron}(a) and (b), diffraction peaks (100) and (102) show transitions at 130 and 440 K respectively. 
 These transitions do not occur in x-ray and electron diffraction,\cite{Wang2013,Wang2014} indicating that the magnetic diffraction contributes significantly in the (100) and (102) peak intensities.
 In particular,the 130 K is also the critical temperature of the weak ferromagnetism $T_W$ in h-LuFeO$_3$ [see Fig. \ref{fig_mag_LFO_Neutron}(c)], indicating a transition of spin structure at $T_W$.
 In contrast, the neutron diffraction intensities do not change significantly for (300) peaks in the temperature range 4 K$<T<$450 K, which is similar to the behavior of the peaks in electron and x-ray diffractions (Fig. \ref{fig_FE_RHEED}). 
 Therefore, the diffraction intensity of the (300) peak is dominated by the nuclear contributions.

 The intensity of the magnetic diffraction of a neutron beam follows:\cite{Lovesey1984}
\begin{equation}
I \propto |\sum \limits_{i} p_i \vec{q}_i  e^{i 2 \pi \vec{h} \cdot \vec{r}_i}|^2,
\end{equation}
where $\vec{q}_i=\vec{h}(\vec{h}\cdot\vec{m}_i)-\vec{m}_i$ , $p_i$  is the isotope specific factor for each site, $\vec{m}_i$ is the unit vector of the magnetic moment and $\vec{h}_i$ is the unit vector perpendicular to the atomic planes involved in the diffraction. 
 Because the factor $\vec{q}_i$ that depends on the orientations of the magnetic moments, does not play a role in the nuclear diffraction, new selection rules are generated. 
 In short, only $\Gamma_1$ and $\Gamma_3$ contributes to the (100) magnetic diffraction intensities. 
 The contribution to diffraction intensities from the Fe$^{3+}$ sites of the $z$=1/2 and $z$=0 layers are cancelled out for $\Gamma_2$ and $\Gamma_4$ magnetic structures.
 This cancellation does not occur for (102) magnetic diffraction because the factors introduced by the different $z$ positions;
 for the (102) peak, the diffraction intensity is non-zero for all the spin structures $\Gamma_1$ through $\Gamma_4$.

 Therefore, the combination of the significant intensity of the (100) neutron diffraction peak and the weak ferromagnetism below $T_W$ indicates that the magnetic structure in h-LuFeO$_3$ is a mixture of $\Gamma_1$ and $\Gamma_2$.
 Above $T_W$, the (100) peak diminishes while (102) peak pesists up to 440 K; this suggests a antiferromagnetic order between 130 K and 440 K, most probably with a $\Gamma_4$ magnetic structure.\cite{Wang2013}
 The coexistence of the antiferromagnetic order and ferroelectricity above room temperature in h-LuFeO$_3$, makes this material room-temperature multiferroic.
 To date, the high-temperature antiferromagnetic order has not been confirmed using other method of characterization, which may be because of the experimental difficulty in determining antiferromagnetism, particularlly in thin film samples.
 Nevertheless, this high temperature antiferromagnetism is an essential issue in h-RFeO$_3$.
 If the room-temperature antiferromagnetism does exist, then the transition at $T_W$ is a spin reorientation, which may be adjusted by tuning the structure of h-RFeO$_3$; 
 this may lead to the simultaneous ferroelectricity and weak ferromagnetism above room temperature in the single phase h-RFeO$_3$.

\section{Magnetoelectric couplings}
 One of the most interesting topic in h-RFeO$_3$ is the possible coupling between the electric and magnetic degrees of freedom.
 This coupling may be manifested as a change of electric (magnetic) properties in a magnetic (electric) field or at a magnetic (ferroelectric) transition.
 There has been some evidence of magnetoelectric coupling reported in the literature. 
 For example, it was observed that the dielectric constant in h-YbFeO$_3$ is sensitive at a ferroelectric transition temperature.\cite{Jeong2012}
 The optical second harmonic generation increases below the $T_W$.\cite{Iida2012}
 More importantly, the desirable property of switching the spontaneous magnetization in h-RFeO$_3$ using an electric field has been predicted theoretically.\cite{Das2014}
 Due to the improper nature of both ferroelectricity and ferromagnetism in h-RFeO$_3$, the structural distortion may mediate the coupling between the ferroelectric and magnetic orders;\cite{Wang2014} this may cause a reversal of magnetization of h-RFeO$_3$ in a electric field.\cite{Das2014}

\section{Conclusion}
  In conclusion, h-RFeO$_3$ is an intriguing family of materials in terms of multiferroic properties.
	Despite the similarity with the RMnO$_3$ family, the uniqueness of the coexisting spontaneous electric and magnetic polarization suggests promising application potentials.
	On the other hand, a great deal of investigations still need to be done, because even the fundamental properties, such as structure, ferroelectricity, and magnetism are under debate.
	In addition, the magnetoelectric couplings in h-RFeO$_3$, as extremely appealing properties predicted by theory, are yet to be studied in different aspects.
	We expect to learn more exciting physics from the material family h-RFeO$_3$ in the future.

\section*{Acknowledgments}

X.S.X. acknowledges the support from the Nebraska EPSCoR.

\appendix
\section{Structural Stability in RFeO$_3$}

\label{marker:stability}

\begin{table*}[ht]
\addtolength{\tabcolsep}{-4pt}

\caption{
 Stable structure type of the selected ABO$_{3}$ compounds at ambient temperature and pressure, where A=Sc, Y, La-Lu, and B=Sc-Ni. 
 The radii (in pm) of the trivalent ion are in the parenthesis.
 The atoms are sorted according to their ionic radii.
 The structures are abbreviated: o for orthorhombic, r for rhombohedral, b for bixbiyte, and h for hexagonal.}
\label{tab:ABO3structure}\centering

\begin{tabular}
{@{}ccccccccc@{}}\toprule 
A\textbackslash{}B Site&Sc(74.5)&Ti(67)&Mn(64.5)&Fe(64.5)&V(64)&Cr(61.5)&Co(61)&Ni(60)\\
\colrule
La (103.2)& o{\cite{Geller1957,Liferovich2004}} & o{\cite{Zhou2005}} & o{\cite{Moussa1996,Dabrowski2005}} & o{\cite{Geller1956b}} & o{\cite{Nguyen1995,Miyasaka2003}} & o{\cite{Geller1957,Hashimoto2000}} & r{\cite{Sis1973}} & r{\cite{Munoz1992}}\\
Ce (101)& & o{\cite{Sunstrom1993}} &  & o{\cite{Robbins1969}} & o{\cite{Nguyen1995,Miyasaka2003}} &  &  & \\
Pr(99)& o{\cite{Geller1957,Liferovich2004}} & o{\cite{Zhou2005}} & o{\cite{Dabrowski2005}} & o{\cite{Geller1956b}} & o{\cite{Geller1957,Miyasaka2003}}& o{\cite{Geller1957}} & o{\cite{Alonso2006}} & o{\cite{Lacorre1991,Munoz1992}}\\
Nd (98.3)& o{\cite{Geller1957,Velickov2007,Liferovich2004}} & o{\cite{Zhou2005}} & o{\cite{Dabrowski2005}} & o{\cite{Geller1956b}} & o{\cite{Geller1957,Miyasaka2003}} & o{\cite{Geller1957}} & o{\cite{Demazeau1974}} & o{\cite{Munoz1992}}\\
Sm (95.8)& o{\cite{Velickov2007,Liferovich2004}} & o{\cite{Zhou2005}} & o{\cite{Dabrowski2005}} & o{\cite{Geller1956b}} & o{\cite{Miyasaka2003}} & o{\cite{Geller1957}} & o{\cite{Demazeau1974}} & o{\cite{Munoz1992,Rosenkranz1998}}\\
Eu (94.7)& o{\cite{Kahlenberg2009,Liferovich2004}}  &  & o{\cite{Tadokoro1998,Dabrowski2005}} & o{\cite{Geller1956b}} &  &  & o{\cite{Demazeau1974}} & o{\cite{Rosenkranz1998,Alonso1999}}\\
Gd (93.8)& o{\cite{Geller1957,Velickov2007,Liferovich2004}} & o{\cite{Zhou2005}} & o{\cite{Kimura2005}}  & o{\cite{Geller1956,Geller1956b}} & o{\cite{Geller1957,Miyasaka2003}} & o{\cite{Geller1957}} &  & \\
Tb (92.3)& o{\cite{Velickov2008,Liferovich2004}} & o{\cite{Zhou2005}} & o{\cite{Kimura2003}} & o{\cite{Marezio1970}} & o{\cite{Miyasaka2003}} & o{\cite{Alonso2006}} & o{\cite{Demazeau1974}} & \\
Dy (91.2)& o{\cite{Velickov2007,Liferovich2004}} & o{\cite{Zhou2005}} & o{\cite{Mori2000,Dabrowski2005}} & o{\cite{Marezio1970}} & o{\cite{Miyasaka2003}} & o{\cite{Krynetskii1997}} & o{\cite{Alonso2006}} & \\
Ho (90.1)& o{\cite{Liferovich2004}} & o{\cite{Zhou2005}} & h{\cite{Yakel1963}} & o{\cite{Marezio1970}} & o{\cite{Sikora2007}} & o{\cite{Tiwari2013}} & o{\cite{Alonso2006}} & \\
Y (90)& o{\cite{Geller1957}} & o{\cite{Zhou2005}} & h{\cite{Yakel1963,Munoz2000}} & o{\cite{Geller1956b,Kawano1994}} & o{\cite{Miyasaka2003}} & o{\cite{Geller1957}} &  & \\
Er (89)&  & o{\cite{Zhou2005}} & h{\cite{Yakel1963}}  & o{\cite{Marezio1970}} & o{\cite{Miyasaka2003}} & o{\cite{Kanekot1977}} & o{\cite{Alonso2006}} & \\
Tm (88)&  & o{\cite{Zhou2005}} & h{\cite{Yakel1963}}  & o{\cite{Marezio1970}} &  & o{\cite{Yoshii2012}} & o{\cite{Alonso2006}} & \\
Yb (86.8)&  & o{\cite{Zhou2005}} & h{\cite{Yakel1963,Isobe1991}} & o{\cite{Marezio1970}} & o{\cite{Miyasaka2003}} & o{\cite{Kojima1984}} & o{\cite{Alonso2006}} & \\
Lu (86.1)&  & o{\cite{Zhou2005}} & h{\cite{Yakel1963,Aken2001}}  & o{\cite{Marezio1970}} & o{\cite{Miyasaka2003}} & o{\cite{Ziel1969}} & o{\cite{Alonso2006}} & \\
Sc (74.5)& b{\cite{Geller1967}} & b{\cite{Shafi2012}} & h{\cite{Yakel1963,Munoz2000}} & b{\cite{Li2012}} & b{\cite{Alonso2004}} &  &  & \\ 
\botrule
\end{tabular}
\end{table*}

\begin{figure}[ht]
\centerline{
\includegraphics[width = 2.5 in]{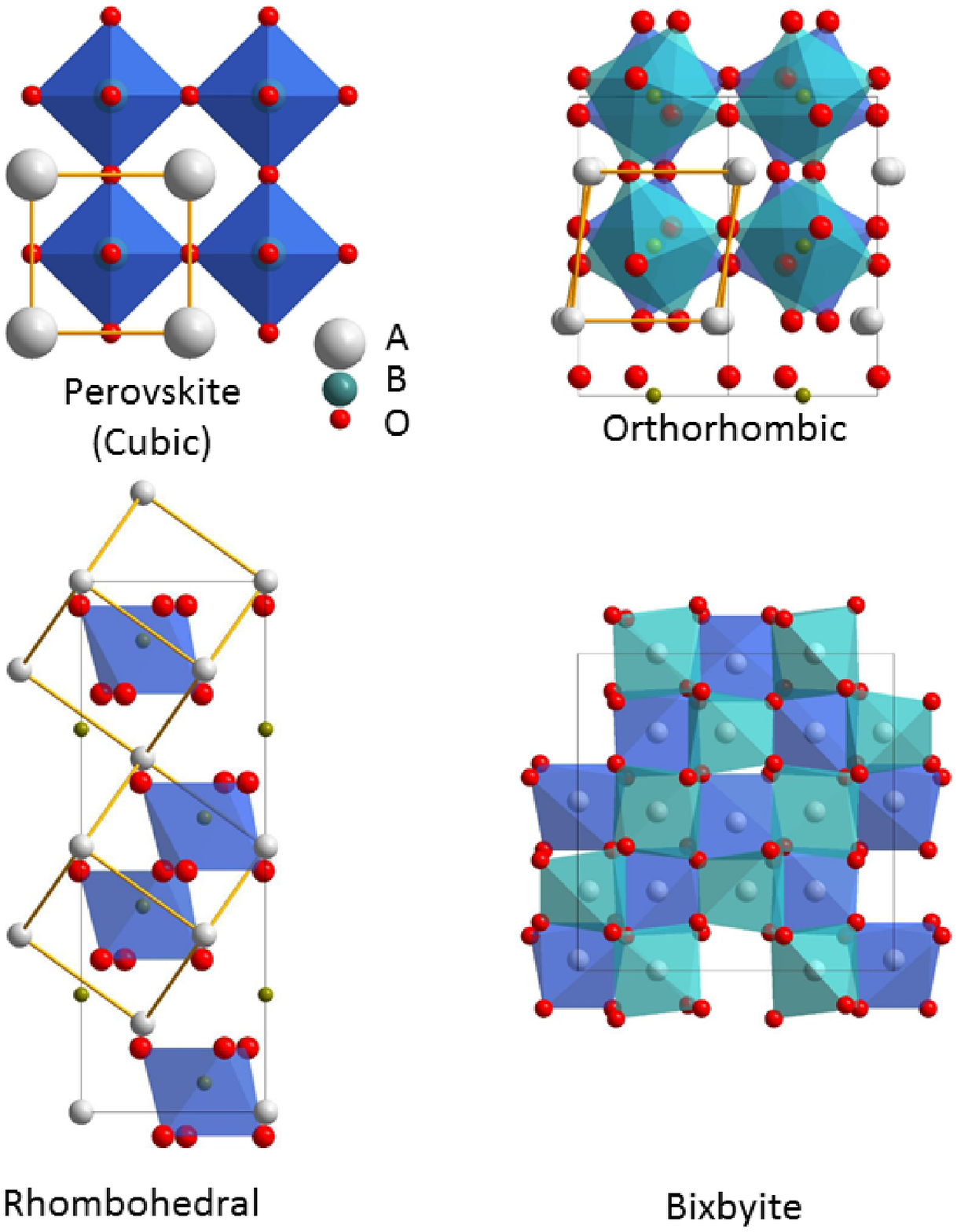}
}
\caption{ (Color online)
	The structures in TABLE \ref{tab:ABO3structure} for ABO$_3$ compounds.
	The perovskite structure contains BO$_6$ octahedra and AO planes.
	The psedo-cubic ABO$_3$ fractions are indicated in the rhombohedral and othorhombic structures.
	The orthorhombic structure can be viewed as the perovskite structure with rotated BO$_6$ octahedra.
	The rhombohedral structure can be viewed as the perovskiste structure distorted along the [111] diagonal direction, in which the A atoms move out of the AO planes.
	In the bixbyite structure, the A and B atoms are not distinguished by the atomic sites they occupy.
}
\label{Fig_ABO3}
\end{figure}

 RFeO$_3$ (R=La-Lu, Y) is known to crystallize in two structure families. 
 The stable structure for bulk stand-alone RFeO$_3$ is orthorhombic (orthoferrite).\cite{Cheremisinoff1990}
 In contrast, the stable structure for manganites are hexagonal for R of small ionic radius (R=Ho-Lu, Y, Sc)\cite{NoteIonicRadius}.
 In order to understand the structural stability of RFeO$_3$, we take an overview of the ABO$_3$ compounds of different A and B sites.
 Table \ref{tab:ABO3structure} displays the stable structure of selected ABO$_3$ compound, where A sites include rare earth, Y, and Sc, and B sites include Sc-Ni.
 The stable structures of most of the compounds are orthorhombic.
 For small A site (Sc), the bixbyite structure is stable.
 When the radius is large for A site and small for B site, the rhombohedral structure becomes stable.
 
 In general, the stability of a crystal structure is related to the ionic radius.
 If the ionic radius of A and B atoms ($r_A$ and $r_B$ respectively) are very different, the stable structure contains two very different sites for metal ions.
 A good example is the perovskite structure (see Fig. \ref{Fig_ABO3} (a)).
 The relation between the ionic radius can be found from the geometry as 

\begin{equation}
t=\frac{r_A+r_O}{\sqrt{2}(r_B+r_O)}=1, 
\label{Eq_tolerance}
\end{equation}

 where $t$ is called tolerance factor and $r_O$ is the ionic radius of O$^{2-}$.\cite{NoteIonicRadius}
 Here the A ions have 12 oxygen neighbor while B ions have only 6 neighbor, corresponding to the large difference between the two kinds of ionic radii .
 On the other hand, perfect satisfaction of Eq. (\ref{Eq_tolerance}) is rare, which is why perfect perovskite structure is rare.
 If $t$ decreases from 1 (the radii of A and B ions become less different), a structural distortion to reduce the coordination of the A ions occurs while keeping the coordination of the B ions as 6;
 this is achieved by either moving A ions out of the AO plane or by rotating the BO$_6$ octahedra.
 As shown in Fig. \ref{Fig_ABO3}, in the distorted structures (rhombohedral or orthorhombic), the A ions move closer to some O$^{2-}$ but away from other O$^{2-}$, reducing the coordination for A ions.
 When the radii of A and B sites are close, the stable structure is bixbyite in which the A and B ions occupy the similar sites.\cite{NoteCorundum}

 On the other hand, the stability of the hexagonal manganites cannot be understood using the argument of atomic radius discussed above.
 More specifically, the structures for RFeO$_3$ and RMnO$_3$ (for R= Ho-Lu, Sc, Y) are very different, while the radii of Fe$^{3+}$ and Mn$^{3+}$ are identical (see Table \ref{tab:ABO3structure}).
 Therefore, the electronic structure must play an important role in the stability of the hexagonal manganites.
 Below, we propose a model to explain why some manganites are stable in hexagonal structure in terms of the electronic structure of Mn$^{3+}$.

\begin{figure}[ht]
\centerline{
\includegraphics[width = 4 in]{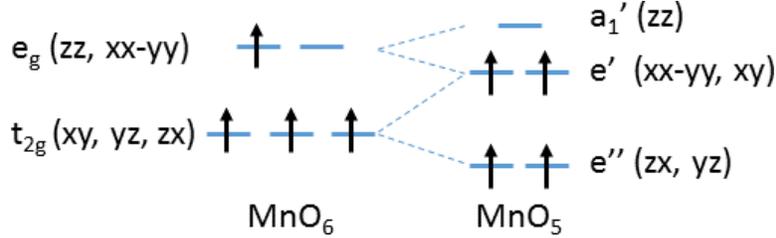}
}
\caption{ (Color online)
	A crystal-field energy model for the stability of hexagonal manganites.
	The electronic configuration of the Mn$^{3+}$ (3d$^4$) in octahedral environment (left) and trigonal bipyramid (right).
	The adoption of the MnO$_5$ local environment may reduce the total energy and stabilized the hexagonal structure.
}
\label{Fig_CFmodel}
\end{figure}

 Fig. \ref{Fig_CFmodel} displays the electronic configuration in the local environment of orthorhombic and hexagonal structures respectively.
 In the orthorhombic structure, the energy level of the 3d electrons in the 6-coordinated Mn$^{3+}$ is split into $t_{2g}$ and $e_g$ levels.
 Only one $e_{g}$ level is occupied while all the $t_{2g}$ levels are singly occupied in Mn$^{3+}$ with a 3d$^4$ configuration, generating a degeneracy.
 In the hexagonal structure, the energy level of the 3d electrons in the 5-coordinated Mn$^{3+}$ is split into $e''$, $e'$, and $a_1'$ levels.
 There is no electronic degeneracy for Mn$^{3+}$ in this case because the highest level $a_1'$ is not degenerate.
 According to the evolution of the 5 3d orbitals (xx-yy, zz, xy, yz, zx), the electronic configuration can be treated as the splitting of $t_{2g}$ and $e_g$ levels due to the MnO$_6$ to MnO$_5$ distortion.
 If we assume that the distortion does not change the total energy of the electronic levels (i.e. $E_{xx-yy}$+$E_{zz}$ and $E_{xy}$+$E_{yz}$+$E_{zx}$ are both conserved), the total energy of the electronic configuration in MnO$_5$ is lower than that in MnO$_6$.
 This scenario is only true for Mn$^{3+}$ with 3d$^4$ configuration because of the electronic degeneracy.
 In contrast, for other trivalent ions (Sc to Ni), the hexagonal structure will not be stable, because the 5-coordinated local environment does not reduce the energy.
 At the same time, the distortion to hexagonal structure must decrease the ionic bonding energy because of the reduced coordination.
 The competition between the two effects stabilized the hexagonal structure for small A ions (Ho-Lu, Sc, Y).
\eject


\begin{references}
\bibitem{NoteIonicRadius} The ionic radius in an oxide depends on the coordination (number of oxygen neighbors) and valence. In this paper, we assume 3+ for the valence and 6 oxygen neighbors. The radius of 6-coordinate O$^{2-}$ is 140 pm.
\bibitem{NoteCorundum} When the raius of the A and B ions are small, a corundum-like structure will be stable (e.g. Fe$_2$O$_3$ and FeTiO$_3$).
\end{references}
\end{document}